\documentclass[
  ,draft            % use draft while you are working on the paper
  ]
  {aipproc}

\layoutstyle{6x9}

\begin{document}

\title{A Decade of Coronagraphic and
  Spectroscopic Studies of CME-Driven Shocks}  

\classification{96.60.Ph, 96.50.Fm, 52.35.Tc, 95.55.Fw }
\keywords      {Coronal Mass Ejections, Interplanetary shocks, Shocks
  in Plasma,Coronagraphs} 

\author{Angelos Vourlidas}{ 
    address={Space Sciences Division, Naval Research Laboratory, Washington DC, USA},
 %   email={vourlidas@nrl.navy.mil}
}
 \author{Alessandro Bemporad}{
   address={INAF-Osservatorio Astronomico di Torino, Pino Torinese, Italy}
 }

\begin{abstract}
  Shocks driven by Coronal Mass Ejections (CMEs) are primary agents of
  space weather. They can accelerate particles to high energies and
  can compress the magnetosphere thus setting in motion geomagnetic
  storms.  For many years, these shocks were studied only in-situ when
  they crossed over spacecraft or remotely through their radio
  emission spectra. Neither of these two methods provides information
  on the spatial structure of the shock nor on its relationship to its
  driver, the CME. In the last decade, we have been able to not only
  image shocks with coronagraphs but also measure their properties
  remotely through the use of spectroscopic and image analysis
  methods. Thanks to instrumentation on \textit{STEREO\/} and
  \textit{SOHO\/} we can now image shocks (and waves) from the low
  corona, through the inner heliosphere, to Earth.  Here, we review
  the progress made in imaging and analyzing CME-driven shocks and
  show that joint coronagraphic and spectrscopic observations are our
  best means to understand shock physics close to the Sun.

\end{abstract}

\maketitle

%%%%%%%%%%%%%%%%%%%%%%%%%%%%%%%%%%%%%%%%%%%%
%% MAINMATTER
%%%%%%%%%%%%%%%%%%%%%%%%%%%%%%%%%%%%%%%%%%%%
\section{The Study of CME-driven Shocks Near the Sun}
Space-based coronagraphs have observed thousands of Coronal Mass
Ejections (CMEs) since their discovery in the early 1970s. The
observations show that hundreds of CMEs can attain speeds of 1000
km/sec or more within a few R$_\odot$ which exceed the local
Alfv\'{e}n speed at those heights. Hence, fast CMEs can drive shocks
which are detected sometimes in the radio wavelengths as drifting
type-II bursts \cite{gopal05}. Such indirect observations, however,
provide only limited information about the shock; namely, the driver
speed and, the onset and duration of the radio emission.

Fortunately, direct imaging and spectroscopic measurements of the
density compression at the shock front have been made possible in the
last ten years thanks to the LASCO coronagraphs and UVCS spectrometer,
both instruments aboard the Solar Heliospheric Observatory
(\textit{SOHO\/}). As we will explain, the combination of white light
and UV observations can provide the full range of physical parameters
(density, temperature, magnetic field) across the shock in addition to
spatial information about the shock shape and its relation to the
driving CME. Now, the physics of coronal shocks
can be studied in detail. The off-limb spectroscopy and
coronagraphy are, and will be, our only means for studying shocks
where they can accelerate particles with the highest energies ($<10$
R$_\odot$). Even the forthcoming \textit{Solar Probe Plus\/} mission
will only reach to 9.5 R$_\odot$. 

\section{Coronagraphic Imaging of Shocks}
   
\begin{figure}
 \includegraphics[height=.5\textheight]{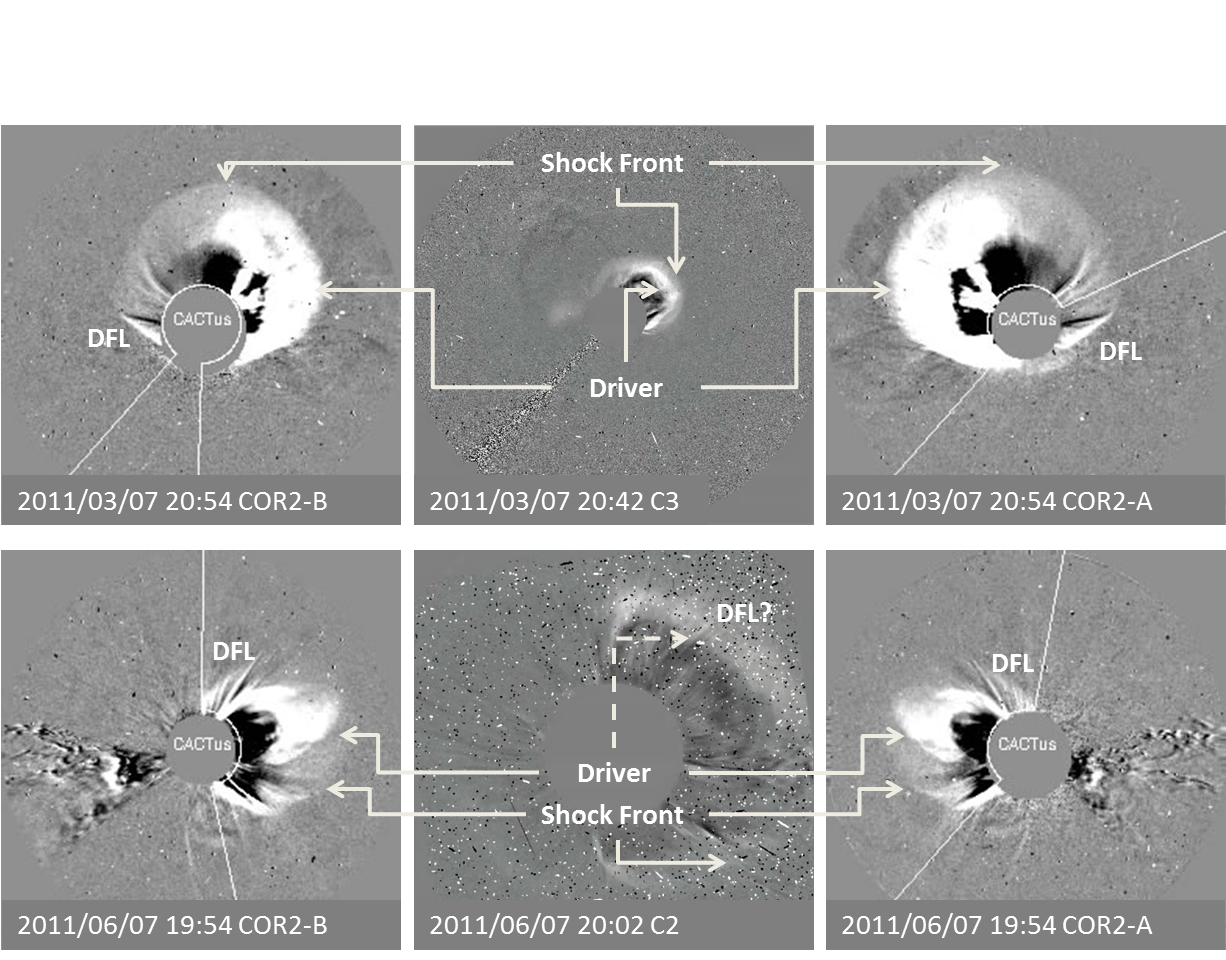}
  \caption{Multipoint coronagraph observations of CME-driven shocks. The times and telescopes are shown on the figure. The angular distance from Earth for COR2-A (COR2-B) are $88^\circ$ ($95^\circ$) on 03/07 and $95^\circ$ ($93^\circ$) on 06/07, respectively. Several streamers are clearly deflected by the shock. Some examples are denoted with 'DFL'. For the June 6 CME, the driver cannot be readily distinguished from the deflected streamers. 
}\label{fig:secchi}
\end{figure}

\begin{figure}
 \includegraphics[height=.5\textheight]{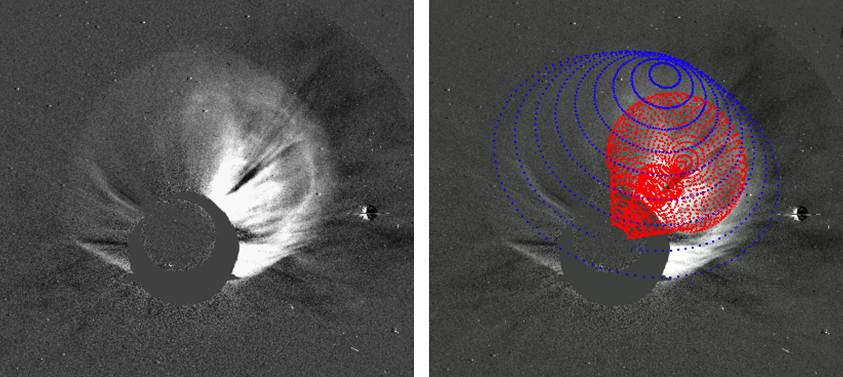}
  \caption{Example of a 3D fitting of the shock and its CME driver for the event on March 7, 2011 using the forward modeling method in \cite{ontiveros09, thernisien09}. 
}\label{fig:shock}
\end{figure}

Because we have reviewed many aspects of coronagraphic imaging of CME
shocks in a recent volume of this conference series \citep{vou09}, we
present here only a brief overview of the progress made over the last
two years. The early work on shock detection \cite{vou03, ontiveros09,
  vou09} has motivated an ever expanding number of detections of
CME-driven shocks \cite{gopal09, temmer11, rouillard11,
  maloney11}. These works are based mostly on observations by the
SECCHI instruments which have higher sensitivites than the LASCO
coronagraphs. The white light shocks can now be routinely detected in
the inner corona, $\sim1.55$ R$_\odot$ \cite{gopal09}, and can be
followed out to at least 0.5 AU \cite{maloney11}. The SECCHI and radio
observations show that shocks tend to form at 1.5 R$_\odot$ and can
accelerate electrons out to 4 $R_\odot$ \cite{gopal09}. The standoff
distance between the shock and the driver appears to be 1-2 R$_\odot$
within the coronagraph field of view \cite{gopal09, maloney11,
  gopal11}. Clear white light shock signatures were rare during
2007-2010 due to the prolonged activity minimum but there have been
several fast events since February 2011. We present two recent
examples in Figs.~\ref{fig:secchi}-\ref{fig:shock}. The three viewpoints afforded by
the SECCHI and LASCO telescopes easily reveal the global reach of
CME-driven shocks in the middle corona.  Measurements of the shock
longitudinal extent allow us to localize with better precision the
injection of Solar Energetic Particles (SEPs) along magnetic field
lines \cite{rouillard11} and connect it to the CME source region
\cite{temmer11}.

\section{Spectroscopic Measurements of Shocks}

Shocks can be identified in UV spectra by means of heating, density compression and high speeds associated with them. The first spectral detection of a CME driven shock was made by UVCS in 2000 \cite{raymond00}. The authors measured enhancements in the O$^{5+}$ and Si$^{11+}$ that were consistent with an MHD shock. More than a dozen of shocks have been identified since \citep{mancuso02,ciaravella05,ciaravella06,mancuso08,mancuso09,bemporad10,mancuso11}. A review of plasma physical parameters derived by UVCS across shocks is given in Table~\ref{tab:a}. The plasma heating and compression across the shock lead typically to strong brightening of the Si$^{11+}$ (Si~{\sc xii} $\lambda$521) line, brightening or dimming of the O$^{5+}$ (O~{\sc vi} $\lambda\lambda$1032-1037) lines and dimming of the H~{\sc i} Ly$\alpha$  $\lambda$1216 line. These differences are related to the different physical processes involved in the atomic excitations, because coronal ions are mainly excited by both collisions with electrons (collisional excitation) and/or by resonant scattering of photons emitted from the underlying layers of the solar atmosphere (radiative excitation); the latter process is also related to the Doppler dimming effect \citep[see][]{noci1987}.

\begin{figure}
  \includegraphics[height=.34\textheight]{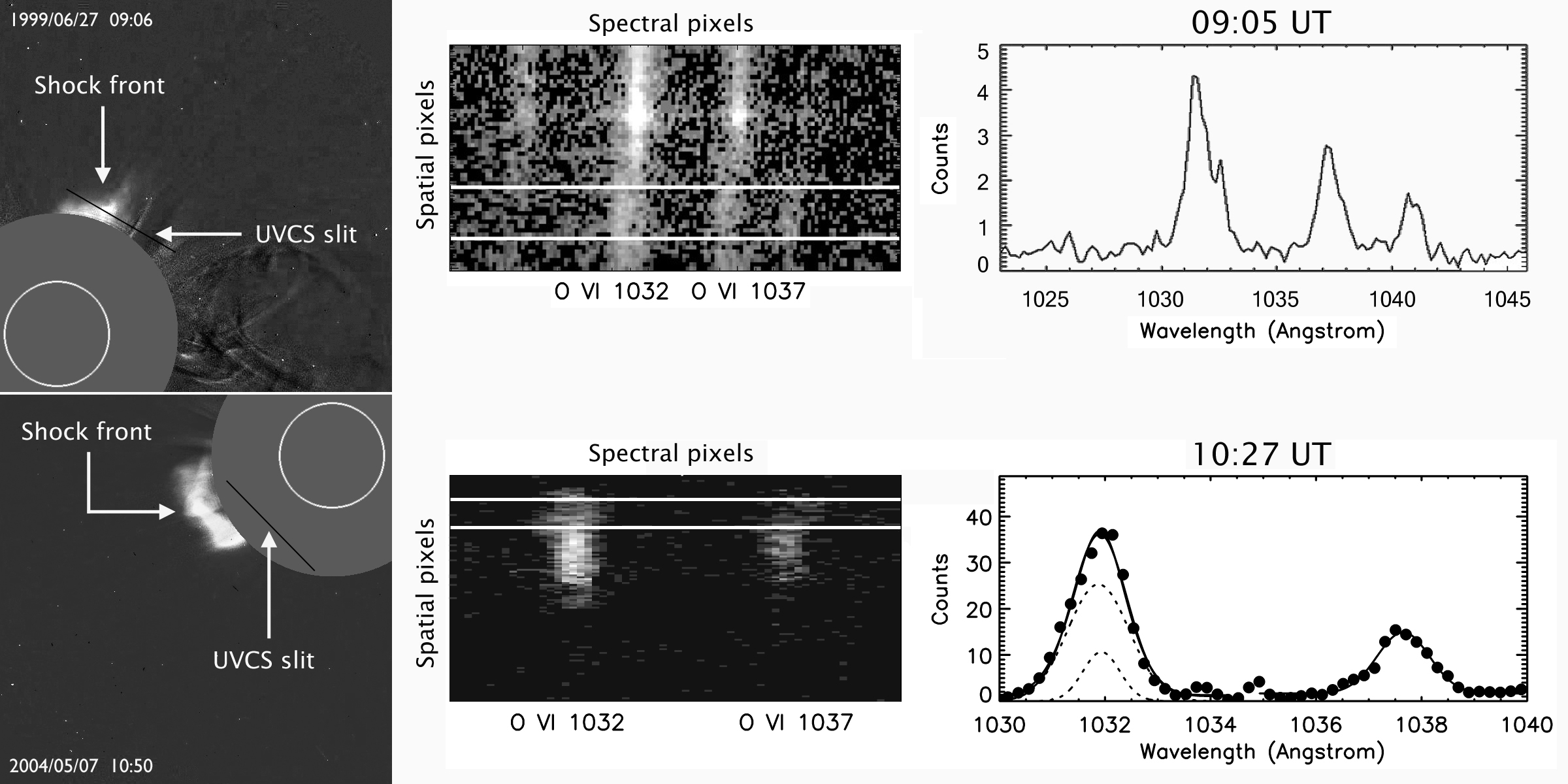}
  \caption{Two examples of shock detections by UVCS. Top left: LASCO/C2 running difference image showing the June 27, 1999 event and the location of the UVCS slit (solid black line) during the observations. Top middle: O~{\sc vi} $\lambda\lambda$1032-1037 intensities observed during the transit of the CME-driven shock. Top right: line profile broadenings observed at the shock transit \citep[average over the spatial region shown in the middle panel by the two solid white lines;][]{raouafi04}. Bottom: same as top panels, but for the May 7, 2004 event \citep[][]{mancuso11}.}\label{fig:uvcsshock}
\end{figure}
The most evident signature of a shock in EUV spectra is a broadening of the O~{\sc vi} $\lambda\lambda$1032-1037 line profiles in close correspondence to the shock transit, as derived from shock speeds either measured in the LASCO coronagraph images or derived from the frequency slope of accompanying type-II radio bursts (Fig.~\ref{fig:uvcsshock}). Spectral signatures of the shocks have been detected from about 1.7 R$_\odot$ up to 4.3 R$_\odot$ (Table~\ref{tab:a}). The observed broadening is usually reproduced with superposition of a narrower component due to emissison from coronal plasma located along the line of sight (LOS), and a broader component due to the shock emission (e.g., Fig.~\ref{fig:uvcsshock} bottom right panel). The latter is often red- or blue-Doppler shifted by up to 0.4\AA\ with respect to the coronal emission because of the LOS component of shock velocity, resulting in asymmetric observed profiles \cite[see e.g.,][and  Fig.~\ref{fig:uvcsshock}, top right panel]{raouafi04,bemporad10}. FWHMs of broad components typically correspond to post-shock O$^{5+}$ kinetic temperatures $T_k \sim 10^8$K.
\begin{table}
\begin{tabular}{lrrrrrr}
\hline
\tablehead{1}{c}{b}{Date}    & 
\tablehead{1}{c}{b}{Height\\(R$_\odot$)}  & 
\tablehead{1}{c}{b}{Speed\\(km s$^{-1}$)}   &
\tablehead{1}{c}{b}{Density\\($10^6$ cm$^{-3}$)} &
\tablehead{1}{c}{b}{Log(T$_k$)}  &
\tablehead{1}{c}{b}{X}       &
\tablehead{1}{c}{b}{Reference} \\
\hline
06/11/98 & 1.75 & 1200 & 1        & 8.7 & 1.8     &\cite{raymond00}\\
06/27/99 & 2.55 & 1200 & ---  & <8.2& --- &\cite{raouafi04}\\
03/03/00 & 1.70 & 1100 & 10       & 8.2 & 1.8     &\cite{mancuso02}\\
06/28/00 & 2.32 & 1400 & 2        & 8.1 &---  &\cite{ciaravella05}\\
07/03/02 & 1.63 & 1700 & 5        & 8.0 & 2.2     &\cite{mancuso08}\\
22/03/02 & 4.30 & 1460 & 0.011     & 7.3 & 2.1     &\cite{bemporad10}\\
07/05/04 & 1.86 & 690 & 5        & <7.0 & ---     &\cite{mancuso11}\\
\hline
\end{tabular}
\caption{CME-driven Shock Parameters Derived from UVCS data}
\label{tab:a}
\end{table}
Broadenings observed in spectral lines emitted by different ions/elements are representative of heatings occurring for different charge-to-mass ratios, thus providing important information on physical mechanisms occurring across the shock surface. For instance, \cite{mancuso02} pointed out that the observed O~{\sc vi} $\lambda$1032 line broadening can be explained by the mechanism proposed by \cite{lee86}, where the heating at a quasi-perpendicular shock is due to the nondeflection of ions at the shock ramp. For another event, \cite{mancuso08} concluded that observed deficiency of O$^{5+}$ ion heating might be attributable to different local plasma conditions, such as higher values of the $\beta$ parameter (i.e. thermal over magnetic plasma pressure) in the corona crossed by the shock. This would be in agreement with \cite{korreck07}, who found that, for supercritical interplanetary shocks, the heating of ions tends to decrease for increasing values of $\beta$. 

Nevertheless, the interpretation is not always straightforward because the line broadening may also be due to non-thermal effects from ions energized at the shock front \citep{kahler99} and to the expansion of the CME front. For instance, after studying 22 full and partial-halo CMEs observed by UVCS, \cite{ciaravella06} concluded that, by assuming hemispherical shapes for the shock fronts, the O~{\sc vi} $\lambda$1032 line broadenings in 3 events can be entirely ascribed to the LOS component of the bulk expansion, while this explanation was inadequate in 5 events. A correct estimate of the post-shock kinetic temperatures requires a subtraction of the line broadening due to the plasma expansion along the LOS.

\begin{table}
\begin{tabular}{ccccc}
\hline
\tablehead{1}{c}{b}{}    & 
\tablehead{1}{c}{b}{T$_e$ ($10^6$ K)}  & 
\tablehead{1}{c}{b}{n$_e$ (10$^{-4}$ cm$^{-3}$)}   &
\tablehead{1}{c}{b}{v (km s$^{-1}$)} &
\tablehead{1}{c}{b}{B (mG)} \\
\hline
upstream 	& 0.23	& 1.1	& 100	& 19\\
downstream 	& 1.9	& 2.3	& 424	& 37\\
\hline
\end{tabular}
\caption{Physical Parameter at a CME-driven Shock derived by Bemporad \& Mancuso (2010)}
\label{tab:b}
\end{table}

There is a high degree of complementarity between the coronagraphic
(spatial distribution, speed of driver and shock) and spectroscopic
measurements (pre- and post-shock plasma parameters). Indeed, combined
analyses of LASCO and UVCS observations can obtain the full range of
plasma parameters upstream and downstream of the shock, including the
magnetic field in both regimes. The MHD Rankine-Hugoniot jump
relations for an oblique shock provide not only the post-shock plasma
parameters, but also the pre- and post-shock magnetic field strength
when the pre-shock coronal plasma parameters, with the exception of
the magnetic field, can be determined by spectroscopic measurements
(i.e., electron density and temperature, pre-shock and shock velocity)
and the density compression ratio across the shock is determined by
coronagraphic measurements. Such results derived by \cite{bemporad10}
are summarized in Table \ref{tab:b}. Interestingly, the magnetic and
kinetic energy density increases are nearly equal (as expected for
equipartition of energy), and are both more than two times larger than
the thermal energy increase. This trailblazing study demonstrates the
power behing off-limb spectroscopic observations when combined with
coronagraphic CME images.

\section{Conclusions}
In the short space of this article, we tried to demonstrate the ability of coronagraphic and spectroscopic instruments to remotely detect coronal shocks and measure their properties. Despite a slow start during the 2000s, shocks are now detected and analyzed with increasing frequency.

Coronagraphs, especially the simultaneous observations from SECCHI and
LASCO, are capable of deriving the 3D structure of the shock, its
standoff distance from the driver, its speed and even the density
compression ratio across the shock boundary. But they do not measure
magnetic field or temperature conditions. Hence, coronagraphs cannot
directly discriminate between a shock or a fast mode wave since the
observed density compression could be the same. However, shocks can
now be followed to at least 0.5 AU and possibly to Earth thanks to
recent imag processing developments \cite{deforest11}. More details
can be found in \cite{vou09}.

The spectroscopic observations have both advantages and disadvantages
compared to the obsevations in the radio or in the visible. On one
hand, the narrow slits provide only one-dimensional spatial
information and the results are quite sensitive to
the orientation of the slit relative to the shock front. Measurements
of very fast shocks may also be compromised due to the relatively long
integrations for the UVCS spectra ($\sim 100$ s). Spectroscopic detection of coronal shocks occurs only by chance,
because the field of view of the 40'' long UVCS slit covers only a
limited latitudinal region (e.g. 75$^\circ$ when the slit is
positioned at 1.7 R$_\odot$). On the other hand, spectral measurements
are the only means to study the heating of individual ion species and
electrons in the collisionless coronal shocks. Measurement of physical
properties in CME-driven shocks and their correlations with SEP events
will be very useful in establishing the necessary conditions for
shocks to accelerate ambient solar wind or suprathermal ions to
SEPs. Fortunately, off-limb spectroscopic capability will be offered
by the forthcoming \textit{Solar Orbiter} mission with the METIS
\citep[Multi-Element Telescope for Imaging and Spectroscopy;
see][]{antonucci11} instrument, that will acquire simultaneously
 off-limb spectra at three heights, thus helping us to disentangle spatial and temporal
evolution of plasma parameters across shocks.

%%%%%%%%%%%%%%%%%%%%%%%%%%%%%%%%%%%%%%%%%%%%%%%%
%% BACKMATTER
%%%%%%%%%%%%%%%%%%%%%%%%%%%%%%%%%%%%%%%%%%%%%%%%

\begin{theacknowledgments}
  Part of this work was funded by various NASA grants. SOHO
  is an international collaboration between NASA and ESA. LASCO was
  constructed by a consortium of institutions: NRL (Washington, DC,
  USA), MPS (Katlenburg- Lindau, Germany), LAM (Marseille, France) and
  Univ.of Birmingham (Birmingham, UK). The SECCHI data are produced by
  an international consortium of the NRL, LMSAL and NASA GSFC (USA),
  RAL and Univ. Bham (UK), MPS (Germany), CSL (Belgium), IOTA and IAS
  (France).

\end{theacknowledgments}

\bibliographystyle{aipproc}   % if natbib is available

\end{document}